\documentclass[12pt]{article}

\input epsf
\def\ba{\begin{eqnarray}}
\def\ea{\end{eqnarray}}
\def\beq{\begin{eqnarray}}
\def\eeq{\end{eqnarray}}

\def\mpl{M_{\rm Pl}}

\def\E{\mathcal{E}}

\def\p{{\cal P}}

\def\L*{{\cal L}_*}
\def\L{\mathcal{L}}
\def\({\left(}
\def\){\right)}
\def\ie{{\it i.e. }}
\def\nn{\nonumber}
\def\p{\partial}
\def\mn{_{\mu \nu}}
\def\lsim{\mathrel{\rlap{\lower3pt\hbox{\hskip0pt$\sim$}}
     \raise1pt\hbox{$<$}}}         %less than or approx. symbol
\def\gsim{\mathrel{\rlap{\lower4pt\hbox{\hskip1pt$\sim$}}
     \raise1pt\hbox{$>$}}}         %greater than or approx. symbol

\usepackage{amsmath}
\usepackage{amsfonts}

\begin{document}

\begin{titlepage}

\begin{flushright}
{NYU-TH-06/13/10}

\today
\end{flushright}
\vskip 0.9cm

\centerline{\Large \bf Generalization of the Fierz-Pauli Action}
\vskip 0.7cm
\centerline{\large Claudia de Rham\,$^1$ and Gregory Gabadadze\,$^2$}
\vskip 0.3cm
\centerline{\em $^1$\,D\'epartment de Physique  Th\'eorique, Universit\'e
de  Gen\`eve,}
\centerline{\em 24 Quai E. Ansermet, CH-1211,  Gen\`eve, Switzerland}

\centerline{\em $^2$\,Center for Cosmology and Particle Physics,
Department of Physics, }
\centerline{\em New York University, New York,
NY, 10003, USA}

\vskip 1.9cm

\begin{abstract}

We consider the Lagrangian of gravity covariantly amended by the
mass and polynomial interaction terms with arbitrary coefficients,
and reinvestigate the consistency of such a theory in the
decoupling limit, up to the fifth order in the nonlinearities.
We calculate explicitly the self-interactions of the helicity-0 mode,
as well as the nonlinear mixing between the helicity-0 and -2 modes.
We show  that ghost-like  pathologies in these interactions disappear
for special choices of the polynomial interactions, and argue that this
result remains true to  all orders in the decoupling limit. Moreover,
we show that the linear,  and some of the nonlinear mixing terms
between the  helicity-0 and -2 modes can be absorbed by a local change
of variables,  which then naturally generates the cubic,  quartic,  and
quintic Galileon interactions, introduced in a different context. We
also point out that the mixing between the helicity-0 and 2 modes can
be at most quartic in the decoupling limit. Finally, we discuss the
implications of our findings  for the consistency of the effective field
theory away from the decoupling limit, and for the Boulware-Deser problem.

\end{abstract}

%\vspace{3cm}

\end{titlepage}

\newpage

\section{Introduction and summary}

In this work we study the covariant polynomial potential of a  relativistic
and symmetric rank-2 tensor field living in four-dimensional flat space-time.

We start with the mass term in the potential. Poincar\'e symmetry in four dimensions imposes that any massive spin-2 state has to have five physical degrees of freedom -- namely, the helicity-$\pm2$, helicity-$\pm1$, and
helicity-$0$ modes. The quadratic potential  that describes these degrees of freedom is that of Fierz and Pauli (FP),  \cite{FP}. The latter is  known  to be the  unique  ghost-free and tachyon-free mass term for the spin-2 state \cite{Nieu}.

No matter how small the graviton mass is in the FP theory, the helicity-0 state couples to the trace of the matter stress-tensor  with the same  strength as the helicity-$2$ does \cite{vDVZ}. This discontinuity  would  rule out, on simple observational
grounds, the FP mass term for gravity.

As argued first  by Vainshtein, the discontinuity problem  can be cured by the nonlinear interactions which would become comparable to the linear terms already for very weak fields \cite{Arkady}.  Then, the  non-linearities could give rise to the screening of the helicity-0  mode at observable scales,  rendering the theory compatible with the known empirical data \cite{Arkady,DDGV}.

However, the very  same non-linearities that cure the discontinuity problem typically give rise to a ghost in massive gravity, \cite{BD}. This ghost, sometimes referred to as the Boulware-Deser (BD) mode, emerges as a sixth degree of freedom, that is infinitely heavy on a flat background, but becomes light on any reasonable nontrivial background  ({\it e.g.}, on a cosmological background \cite{GGruzinov}, or on the weak background of a lump of static matter \cite{AGS,Creminelli,DeffayetRombouts}). It is straightforward to see this ghost  in the so-called decoupling limit  \cite{AGS}, in  which the dynamics of the helicity-0 mode can be made manifest.  Then, the sixth degree of freedom ends up being related to the nonlinear interactions of the helicity-0 mode \cite{AGS,Creminelli,DeffayetRombouts}\footnote{Notice also that the discontinuity is absent when a small cosmological
constant is included  before sending the mass of the graviton to zero
\cite {Kogan,Porrati}. Doing so in
de Sitter space, however, one passes  through
the parameter region where helicity-0 becomes a ghost
\cite {Higuchi}, while the anti de Sitter case
is ghost-free \cite {Kogan,Porrati}.}.

The obvious question to ask is then whether there exists a nonlinear model that exhibits the Vainshtein mechanism, but without the ghost mode. This  question  was raised  in Ref.~\cite{AGS}, and studied  in detail in Ref.~\cite{Creminelli}. The latter
work argued that  at the cubic order the ghost can be avoided by tuning the coefficients of the quadratic and cubic order terms. Recently, the cubic  terms were calculated in a nonlinear massive  spin-2  theory of  Refs.~\cite{GG,Claudia}, where it was  shown that the necessary tuning is in fact automatic in this model, and the theory is ghost-free to that order \cite{cubic}!

In the present work we focus instead on addressing this question at higher orders,
and in a model-independent framework. We therefore allow for arbitrary nonlinearities in the potential up to the quintic order, but restrict ourselves to considerations in  the decoupling limit only.

Our result clashes with one of the conclusions of Ref.~\cite{Creminelli} which states that the  quartic interactions in the decoupling limit ineradicably lead to a ghost.  Regretfully,  the decoupling limit Lagrangian obtained in Ref.~\cite{Creminelli} is not reparametrization invariant neither at the cubic nor quartic order, and gives a tensor equation that does not  satisfy the Bianchi identity.  The ghost found in the decoupling limit of Ref.~\cite {Creminelli} is an artifact of these properties.
Hence, we re-investigate this issue in the present work. We find a decoupling limit Lagrangian that is similar to that of Ref.~\cite{Creminelli},  but differs from it  in detail, by coefficients of various tensorial  structures. In particular, due to those coefficients, our Lagrangian is reparametrization invariant, and naturally leads to a tensor equation for which the Bianchi identity is automatically satisfied (as it should be since the helicity-2 mode only mixes linearly in the decoupling limit). Then, not surprisingly,  we arrive to a different conclusion,  that the quartic theory is also ghost-free in the decoupling  limit.
Moreover, we go on one step further and investigate the quintic-order
theory,  which we also show is ghost-free in the decoupling limit.
This also allows us to understand the structure of the interactions to all orders
and to argue that the decoupling limit can be at most quintic order in
interactions (or quartic in the mixing between the helicity-0 and 2 modes)
in the ghost-free theory.

Finally, as a corollary, we  find that the decoupling limit  of the most
general consistent theory of massive gravity  gives rise to the quadratic,
cubic, quartic and quintic Galileon kinetic interactions  introduced in
Ref.~\cite{Nicolis:2008in} in a different context (namely,
as a generalization of  the special cubic term appearing in the
decoupling limit of DGP \cite{Dvali:2000hr} found in Ref.~\cite{Ratt}).
The Galileon interactions share
the important properties of (i) being local, (ii) preserving
the shift and galilean symmetry  in the  field space of the helicity-0 mode
(in particular, in the kinetic and self-interaction terms but not in
interactions with  matter),
(iii) giving rise to equations of motion with a well-defined Cauchy problem.
Since then, the Galileons have developed  their own independent and interesting life
(see, {\it e.g.}, \cite{CedricGalileon,deRham:2010eu}).  We show here that the
Galileons naturally arise in the decoupling limit of a general theory of
massive
gravity. This also helps to prove that upon appropriate choices of the coefficients in the potential,  the decoupling limit of massive gravity is stable, at least up to the quintic order in interactions.

We continue this section with a  discussion  and summary of our main results in more technical terms, before turning to the detailed calculations in the subsequent sections.

\vspace{0.1cm}

In analogy with a massive non-Abelian (Higgs-less) spin-1 \cite{Khriplovich},  the dynamics of the helicity-0 mode,  $\pi$,  can be extracted in a generic theory of gravity with a nonlinear potential by taking the decoupling limit \cite{AGS}
\beq
m \to 0, ~~~\mpl \to \infty, ~~~  {\rm keeping\ } \Lambda_5 \equiv
(m^4 \mpl )^{1/5}
 ~~{\rm fixed}\,.
\label{declim5}
\eeq
Following \cite{AGS}, in a generic case of the nonlinear potential,  the corresponding Lagrangian for the helicity-0 mode reads schematically as follows:
\beq
\mathcal{L}_\pi = {3\over 2} \pi \square \pi +
{(\partial^2 \pi)^3\over \Lambda_5^5}\,.
\label{Lpi5}
\eeq
The cubic  interaction with six derivatives gives rise to a ghost on  locally nontrivial asymptotically-flat  backgrounds ({\it e.g.} on the background of  a local lump of matter).  This could be seen by observing that for $\pi = \pi^{cl} +\delta \pi$, with $\pi^{cl}$ denoting the weak field of a local source, and   $\delta \pi$ its fluctuation, the cubic term in (\ref{Lpi5}) could generate a four-derivative quadratic term for the fluctuations. This leads to  a ghost, which is infinitely heavy on Minkowski space-time, but becomes light enough to be disruptive  once a reasonable local background is considered, see Refs.~\cite{AGS,Creminelli,DeffayetRombouts}.

To avoid pathologies such as in \eqref{Lpi5}, the Fierz-Pauli combination in the  graviton potential should  be pursued further by tuning the coefficients of various higher order terms. This leads to a cancelation of all the terms for $\pi $ that are suppressed by the scales $\Lambda_5$, $\Lambda_4=(m^3 \mpl)^{1/4}$, $\Lambda_{11/3}=(m^{8}\mpl^3)^{1/11}$ etc\ldots for any scale $\Lambda<\Lambda_3=(m^2 \mpl)^{1/3}$, such that only the terms suppressed by the scale $\Lambda_3$  survive. Then,  $\Lambda_3$ is  kept  fixed  in the decoupling limit, and the surviving terms (in addition to the linearized Einstein-Hilbert term) read as follows:
\ba
\Delta \mathcal{L}=h^{\mu\nu}\(X^{(1)}\mn+\frac{1}{\Lambda_3^3}
X^{(2)}\mn+\frac{1}{\Lambda_3^6}X^{(3)}\mn\)\,.
\label{s1}
\ea
Here, $h_{\mu\nu}$ denotes the canonically normalized (rescaled by  $\mpl$)
tensor field perturbation, while  $X^{(1)}\mn,X^{(2)}\mn,$ and  $X^{(3)}\mn$ are
respectively, linear, quadratic and cubic in $\pi$.  Importantly, they are all transverse (for instance, $X^{(1)}\mn  \propto \eta_ {\mu\nu}\square\pi - \partial_\mu \partial_\nu \pi$). Not only do these interactions automatically satisfy the Bianchi identity,  as they should to preserve diffeomorphism invariance, but they are also at most second order in time derivative. Hence, the interactions (\ref{s1}) are  linear in the helicity-2 mode, and unlike the previous results in the literature, present perfectly
consistent terms, at least up to the quintic order.

Furthermore, some of the  terms  in (\ref {s1}) can be absorbed  by a  local field redefinition. For instance, the  quadratic term,  $h^{\mu\nu} X^{(1)}\mn$, can be absorbed  by a conformal transformation $h\mn \to h\mn + \eta\mn \pi$.
This shift,  besides removing the above mixing, generates terms of the form $ \pi X^{(2)} $ and $ \pi X^{(3)} $, which coincide,  up to a total derivative, with the cubic and quartic Galileon terms \cite{Nicolis:2008in}.  Further  diagonalization of the cubic mixing term, $h^{\mu\nu} X^{(2)}\mn$, also generates the quintic Galileon, hence exhausting all the possible terms that can arise in the Galileon family at arbitrary order.

Moreover, we also point out that if the decoupling limit happens to pick up the scale  $\Lambda_3$ (as opposed to another smaller scale such as $\Lambda_5$, $\Lambda_4$, etc\ldots), the mixing between the  helicity-0 and -2 modes must stop at the quartic order. Therefore,  for  appropriate choices of the interaction coefficients, the decoupling limit at this order is exact! It is the subsequent diagonalization of the  nonlinear terms in the Lagrangian that generates the quintic Galileon.

Finally, the  absence of a ghost in the decoupling limit does not prove the stability of the full theory away from the limit and the Boulware-Deser ghost is still expected to be present in general. However, it at least shows that one has a well-defined and consistent effective field theory below the  scale $\Lambda_3$. Above this scale, the full theory has to be specified. We discuss related issues in section 5. Before that, our work has a two-fold motivation: (i) To establish a consistent effective field theory below $\Lambda_3$ (for the full theory to be viable its decoupling limit should  be ghost-free as a necessary condition). (ii) All the known examples show that the Boulware-Deser ghost, if present in the full theory, does also show up in the decoupling limit. Therefore, it is
encouraging to find no ghosts in this limit.

The paper is organized as follows: In section 2 we summarize  the
formalism used to study the decoupling limit of massive gravity with a
general potential.  We then explicitly compute the decoupling limit Lagrangian
to  the quartic and quintic orders in section 3.
We work with a generic nonlinear completion of the FP gravity for
which the
scale $\Lambda_3^3=\mpl m^2$ is fixed. We argue that the $\pi$ mode
does not decouple from the tensor mode, but that the interactions are
free of any ghost-like
pathologies. In section 4 we give a general framework
for computing the Lagrangian in the decoupling limit, and argue that
in theories which are consistent with the fixed scale  $\Lambda_3$, at most the
quartic order mixing term can be obtained, all the higher order mixing
terms being zero. Moreover, we show in section 5 that upon an appropriate
change of variables we recover the standard Galileon interactions.
Section 6 contains some discussions on open
issues and future directions addressing the consistency of massive gravity
away from  the decoupling limit.

\section{Formalism}

\subsection{Gauge invariant potential for gravity}

\label{GI}
Below  we consider in detail the decoupling limit of a general Lagrangian
of a massive spin-2 field endowed with a potential on Minkowski space-time.
We use the technique developed in Ref. \cite {AGS}. The covariant
Lagrangian with the potential reads as follows:
\ba
{\cal L} =  M^2_{\rm Pl} \sqrt{-g}  R -   \frac{ M^2_{\rm Pl} m^2}{4}
\sqrt{-g} \({U}_2(g,H)+{U}_3(g,H)+{U}_4(g,H)+ {U}_5(g,H)\cdots\) \,,
\label{PF}
\ea
where $U_i$ denotes the interaction term at $i^{\rm th}$ order in $H\mn$,
\ba
\label{PFS}
{U}_2(g,H)&=&H^2_{\mu\nu}-H^2\,,\\
{U}_3(g,H)&=&c_1 H\mn^3+c_2 H H\mn^2+c_3 H^3\label{L3}\,,\\
{U}_4(g,H)&=&d_1 H\mn^4+d_2 H H\mn^3+d_3 H\mn^2H_{\alpha\beta}^2+
d_4 H^2 H\mn^2+d_5 H^4\label{L4}\,,\\
{U}_5(g,H)&=&f_1 H\mn^5+f_2 H H\mn^4+f_3 H^2 H\mn^3+
f_4 H_{\alpha\beta}^2 H\mn^3 \\
&+&f_5H (H\mn^2)^2+f_6H^3 H\mn^2+f_7 H^5 \label{L5}\,.  \nonumber
\ea
Here the index contractions are performed using the inverse metric, so that $H=g^{\mu\nu}H\mn$, $H\mn^2=g^{\mu\nu}g^{\alpha\beta}H_{\mu\alpha}H_{\nu\beta}$, etc\ldots. The coefficients $c_i$, $d_i$ and $f_i$ are a priori arbitrary,  but  will be determined by demanding that no ghosts are present at least up to the quintic order in the decoupling limit.

Finally, the tensor $H\mn$ is related to the metric tensor as follows:
\ba
g\mn&=&\eta\mn+\frac{h\mn}{\mpl}\nn\\
&=&H\mn+\eta_{ab}\partial_\mu \phi^a \partial_\nu \phi^b\,,
\ea
where $a,b =0,1,2,3,$  $\eta_{ab}={\rm diag} (-1,1,1,1)$, and $H\mn$ is a
covariant tensor as long as the four fields  $\phi^a$ transform  as scalars  under  a change of coordinates. Furthermore,  expressing $\phi^a$ in terms of the coordinates $x^\alpha$,
and the field $ \pi^\alpha $ as
$\phi^a= (x^\alpha-\pi^\alpha)\, \delta^a_\alpha$, we obtain
\ba
\label{Hmn}
H\mn=\frac{h\mn}{\mpl}+\partial_\mu \pi_\nu + \partial_\nu \pi_\mu -
\eta_{\alpha\beta}\partial_\mu \pi^\alpha \partial_\nu \pi^\beta\,.
\ea
In (\ref{Hmn}), and in what follows,  we adopt the convention that the
indices on $\pi_\mu$ are raised and lowered with respect to  the Minkowski
metric $\eta\mn$.
Crucially,  the expression for the tensor $H\mn$ in \eqref{Hmn} differs by
a minus sign in front of the last term from the analogous expression
in eq.~(5)
used  in Ref.~\cite{Creminelli}. To emphasize the importance of this sign, we
derive in \ref{AppCreminelli} the decoupling limit using the opposite
sign in (\ref {Hmn}), recover the results  of Ref.~\cite{Creminelli},
and show that the Bianchi identity is then not automatically satisfied,
since the reparametrization invariance is not retained in  the resulting
Lagrangian.

From  (\ref {PF}) it is not immediately clear what is the
scale of the effective field theory represented by this Lagrangian,
{\it i.e.}, what is the energy/momentum scale by which the
higher polynomial interactions would be suppressed as compared
with the leading ones. This will become clear by studying the decoupling
limit of the theory.

In what follows, we focus on the helicity-2 and  helicity-0 modes,
but ignore the vector mode. The latter enters only
quadratically in the decoupling
limit (since  the vector does not couple to a conserved stress-tensor
in the linearized order), and can be   set to zero self-consistently.
This does not prove that  the vector sector is ghost-free,
however, the findings  of Ref.~\cite{cubic} that
the cubic nonlinearities for the vector are completely
harmless due to the $U(1)$ gauge invariance of the resulting terms,
suggest that the vector sector is not going
to reintroduce the BD ghost. Therefore, we use the substitution:
$
\pi_\alpha=\partial_\alpha \pi/ \Lambda_3^3,
$
so that
\ba
H\mn&=&\frac{h\mn}{\mpl}+\frac{2}{\mpl m^2}\Pi\mn-\frac{1}
{M_{\rm Pl}^2m^4}\Pi\mn^2\,,
\ea
where we use the same notation as in \cite{Creminelli},
$\Pi\mn=\p_\mu\p_\nu \pi$ and
$\Pi\mn^2=\eta^{\alpha\beta}\Pi_{\mu\alpha}\Pi_{\beta \nu}$.
Moreover, in what follows  the square brackets $[\ldots]$ will
represent the trace of a tensor contracted using the Minkowski
metric, {\it e.g.} $[\Pi^2]=\Pi^{\mu\nu}\Pi_{\mu\nu}$ and
$[\Pi]^2=\Pi^{\mu}_{\mu} \Pi^{\nu}_{\nu}$.

\subsection{Decoupling scale}

As mentioned in the introduction, the interactions $U_2$ and $U_3$ typically lead to terms of the form $(\partial^2 \pi)^3/(\mpl m^4)$, and the decoupling limit should be taken keeping the scale $\Lambda_5^5=\mpl m^4$ fixed, while $\mpl\to \infty $ and $m\to 0$. However we will show in what follows (see also \cite{cubic})
that for some special values of the coefficients $c_i$, such
interactions cancel (up to a total derivative),
generalizing the FP term to the cubic order. This procedure can be extended
further to an arbitrary order:

At a given order the leading contributions are of the form
\ba
\label{Ln bad}
\mathcal{L}_n\sim \frac{(\partial \partial \pi)^n}{M_{\rm Pl}^{n-2}m^{2(n-1)}}\,,
\ea
then, one chooses the interactions $U_n(H)\sim H^n$ so that the above  terms combine into a total derivative. At each order, there exists a unique total derivative combination $\mathcal{L}_{\rm der}^{(n)}$ that can be written as follows:
\ba
\label{Lder n}
\mathcal{L}_{\rm der}^{(n)}=-\sum_{m=1}^{n}(-1)^m\frac{(n-1)!}{(n-m)!}\,
[\Pi^{m}]\,\mathcal{L}^{(n-m)}_{\rm der}\,,
\ea
with $\mathcal{L}^{(0)}_{\rm der}=1$ and $\mathcal{L}^{(1)}_{\rm der}=[\Pi]$.
Up to the quartic order, the total derivatives are
\ba
\label{L2der}
\mathcal{L}^{(2)}_{\rm der}&=&[\Pi]^2-[\Pi^2]\,,\\
\label{L3der}
\mathcal{L}^{(3)}_{\rm der}&=&[\Pi]^3-3 [\Pi][\Pi^2]+2[\Pi^3]\,,\\
\label{L4der}
\mathcal{L}^{(4)}_{\rm der}&=&[\Pi]^4-6[\Pi^2][\Pi]^2+8[\Pi^3]
[\Pi]+3[\Pi^2]^2-6[\Pi^4]\,.
\ea
Moreover, at higher orders these total derivatives vanish identically, \ie $\mathcal{L}^{(n)}_{\rm der}\equiv 0$,  for any $n\ge 5$. By ensuring that all the leading terms \eqref{Ln bad} take the form of a total derivative \eqref{Lder n}, all the interactions that arise at an energy scale lower than $\Lambda_3$  disappear. Keeping this in mind we will therefore consider below the following
decoupling limit (firts considered in \cite {Ratt} in the context of the
DGP model)
\beq
m \to 0, ~~~\mpl \to \infty, ~~~  {\rm keeping\ } \Lambda_3 \equiv  (m^2 \mpl )^{1/3}
 ~~{\rm fixed}\,.
\label{declim3}
\eeq
Note that the procedure of taking the limit in the present case is well
defined for fields that decay fast enough at spatial infinity. For these
we introduce an  infrared regulator of the theory,
say a large sphere of radius $L \gg 1/m$, and  take the radius
to infinity,  $L\to \infty $, before taking the limit (\ref{declim3}).
This hierarchy of scales enables us to put all the surface
terms to zero before taking the decoupling limit.

Furthermore, as it should be becoming clear from the
above discussions, the scale $\Lambda_3$ will end
up being the effective field theory scale. The higher
interaction terms, both written  or implied
in (\ref {PF}), will be subdominant to the leading
ones  for energy/momentum scales below $\Lambda_3$.

\section{Decoupling limit of massive gravity}

\subsection{Cubic order}
We now explicitly compute the decoupling limit for the
interactions considered in (\ref{PFS}-\ref{L5}),  and thus
generalize the Fierz-Pauli term to higher orders.
In terms of the ``Einstein operator" $\hat \E$ defined for
an arbitrary symmetric field $Z\mn$  as
\ba
\hat \E^{\alpha\beta}_{\mu\nu}Z_{\alpha\beta}=-\frac 12
\(\Box Z\mn-\p_\mu \p_\alpha Z^\alpha_\nu-\p_\nu \p_\alpha Z^\alpha_\mu+
\p_\mu\p_\nu Z^\alpha_\alpha
-\eta\mn \Box Z^\alpha_\alpha+\eta\mn \p_\alpha \p_\beta Z^{\alpha\beta}\),
\ea
the decoupling limit Lagrangian of  massive gravity up to the cubic
order reads as follows
\ba
\mathcal{L}&=&-\frac 12 h^{\mu \nu} \hat \E^{\alpha\beta}_{\mu\nu}h_{\alpha\beta}+
 h^{\mu \nu}X^{(1)}\mn\\
&&-\frac{1}{4\Lambda_5^5}\((8c_1-4)[\Pi^3]+(8c_2+4)[\Pi]
[\Pi^2]+8c_3[\Pi]^3\)\nn
+\frac{1}{\Lambda_3^3}h^{\mu\nu}X^{(2)}\mn\,,
\ea
with
\ba
X^{(1)}\mn&=&[\Pi]\eta\mn-\Pi\mn\,, \label{X1}
%\\
%X^{(2)}\mn&=&-\(3c_1-\frac 52\)\Pi^2\mn-2(1+c_2)[\Pi]\Pi\mn \\
%&+&\(\frac 12 -3c_3\)[\Pi]^2\eta\mn-\(1 +c_2\)[\Pi^2]\eta\mn\nn\,.
\ea
and $X^{(2)}\mn$ quadratic in $\Pi$.
Using the total derivative combination \eqref{L3der}, the
interactions arising at the scale $\Lambda_5$ can be removed by setting
\ba
\label{c1 and c2}
c_1=2c_3+\frac 1 2 \hspace{20pt}{\rm and}\hspace{20pt}c_2=-3c_3-\frac 12\,.
\ea
As a result, we find the following expression for the tensor $X^{(2)}\mn$
\ba
\label{X2}
X^{(2)}\mn=-(6c_3-1)\Big\{(\Pi^2\mn-[\Pi]\Pi\mn)-\frac
12 \([\Pi^2]-[\Pi]^2\)\eta\mn\Big\}\,.
\ea
Notice that both $X^{(1)}\mn$ and  $X^{(2)}\mn$ are automatically conserved, as they should for the reparametrization invariance to be retained and the Bianchi identity to be satisfied.

Moreover, it is straightforward to check that these cubic interactions bear at most two  time derivatives, and are therefore free of any ghost-like pathologies. One should also check that the lapse (which coincides with $h_{00}$ in the decoupling limit) still propagates a constraint,  which is indeed the case here as neither $X^{(1)}_{00}$ nor $X^{(2)}_{00}$ contain  any time derivatives. Furthermore, these cubic interactions with the specific coefficient $c_3=1/4$ have already been discussed in detail in Ref.~\cite{cubic}.

We now apply the same formalism to quartic interactions for which ghost-like
pathologies have been argued to arise inexorably in Ref.~\cite{Creminelli}.

\subsection{Quartic order}
At the quartic order, we find the following interactions
in the decoupling limit:
\ba
\mathcal{L}^{(4)}=
\frac{1}{\Lambda_3^6}h^{\mu\nu}X^{(3)}\mn+
\frac{1}{\Lambda_4^8}\Big\{
(3c_1 -4d_1-\frac{1}{4})[\Pi^4]+(c_2-4d_3+\frac{1}{4})[\Pi^2]^2\\
% &&\hspace{30pt}
+(2c_2-4d_2)[\Pi][\Pi^3]+(3c_3-4d_4)[\Pi^2][\Pi]^2-4d_5[\Pi]^4
\Big\}\nn\,,
\ea
with $\Lambda_4=(\mpl m^3)^{1/4}$ and $X^{(3)}\mn$ cubic in $\Pi$.
%\ba
%X^{(3)}\mn\hspace{-8pt}&=&\hspace{-8pt}(-2+9c_1-8d_1)\Pi^3\mn-(-1-5c_2+6d_2)[\Pi]\Pi^2\mn\\
%\hspace{-8pt}&+&\hspace{-8pt}\nn(1+3c_2-8d_3)[\Pi^2]\Pi\mn+(6c_3-4d_4)[\Pi]^2\Pi\mn+\Big\{-(c_3+8d_5)[\Pi]^3\\
%\hspace{-8pt}&+&\hspace{-8pt}\frac 12(1-2c_1+2c_2-4d_2)[\Pi^3]-\frac 12(1+2c_2-6c_3+8d_4)[\Pi^2][\Pi]\Big\}\eta\mn\nn\,.
%\ea
Here  again the pathological terms arising at the scale $\Lambda_4$
can be removed by using the total derivative combination \eqref{L4der},
and by setting $c_1$ and $c_2$ as in \eqref{c1 and c2},  as well as
\ba
d_1&=&-6d_5+\frac{1}{16}(24c_3+5)\,,\label{d1}\\
d_2&=&8d_5-\frac{1}{4}(6c_3+1)\,,\\
d_3&=&3d_5-\frac{1}{16}(12c_3+1)\,,\\
d_4&=&-6d_5+\frac34 c_3\,.\label{d4}
\ea
Substituting these coefficients in $X^{(3)}\mn$ we obtain the mixing
term between the helicity-0 and 2 modes determined by
\ba
\label{X3}
X^{(3)}\mn=\(c_3+8d_5\)\Big\{6\Pi\mn^3-6[\Pi]\Pi\mn^2+3([\Pi]^2-[\Pi^2])
\Pi\mn\\
-\([\Pi]^3-3[\Pi][\Pi^2]+2[\Pi^3]\)\eta\mn\Big\}\nn\,.
\ea
This expression bears two expected but important features:
\begin{itemize}
\item It is conserved $\partial^\mu X^{(3)}\mn=0$, as it should be for the
reparametrization invariance to be present  and the Bianchi identity
to be automatically satisfied.
\item For $i,j$ space-like indices and $0$ time-like index:
\end{itemize}\vspace{-10pt}
\ba
&& X^{(3)}_{ij} \hspace{15pt} \text{has at most two time derivatives,}\nn \\
&& X^{(3)}_{0i} \hspace{15pt} \text{has at most one time derivative,}\nn \\
&& X^{(3)}_{00} \hspace{15pt} \text{has no time derivatives}.\nn
\ea
These  properties  ensures that no ghost-like pathology arise at the
quartic level in the decoupling limit as long as the interactions come
in with the generalized FP structure set by the coefficients
(\ref{c1 and c2}) and (\ref{d1}-\ref{d4}).

\subsection{Quintic order}
At the fifth order in the decoupling limit, we consider interactions
as given in (\ref {L5}).  The pathological terms that scale as
\ba
\mathcal{L}_{\Pi^5}\sim\frac{1}{M^3_{\rm Pl}m^8}(\p \p \pi)^5\,,
\ea
can be canceled with an appropriate choice of the coefficients $f_1$ to $f_6$:
\ba
\label{fs}
\begin{array}{ccc}
f_1=\frac{7}{32}+\frac{9}{8}c_3-6d_5+24f_7\ ,
& \hspace{20pt}& f_2 = -\frac{5}{32} -\frac{15}{16}c_3+6d_5-30f_7\ ,\\
f_3=\frac38 c_3-5d_5+20 f_7\ ,
& \hspace{20pt}& f_4=-\frac{1}{16}-\frac34 c_3+5d_5-20f_7\ ,\\
f_5=\frac{3}{16} c_3-3d_5+15f_7\ , & \hspace{20pt}& f_6=d_5-10f_7\,.
\end{array}
\ea
As a result,  the quintic interactions in $\pi$ arrange themselves
to form the expression for $\mathcal{L}^{(5)}_{\rm der}$,  as derived
from \eqref{Lder n}
\ba
\mathcal{L}^{(5)}_{\rm der}&=&24[\Pi^5]-30[\Pi][\Pi^4]+20[\Pi^3]([\Pi]^2-[\Pi^2])\\
&& +15[\Pi][\Pi^2]^2-10[\Pi^2][\Pi]^3+[\Pi]^5\nn\equiv 0\,.
\ea
Notice that $\mathcal{L}^{(5)}_{\rm der}$ is not simply a total derivative as
for the previous orders, but instead vanishes identically.
This implies in particular that any limiting Lagrangian of the form
$\mathcal{L}^{(n)}\sim f(\Pi) \mathcal{L}^{(5)}_{\rm der}$, where
$f$ is an analytic function,  gives no dangerous $\pi$
interactions and can be used at  higher orders. Beyond the
quintic order the degrees of freedom in the coefficients
to be tuned should therefore increase, and make it easier to remove
any ghost-like interactions.

With the above  choice of coefficient \eqref{fs}, the only
quintic interaction in the decoupling limit then is
\ba
\mathcal{L}^{(5)}=\frac{1}{\Lambda_3^9}h^{\mu\nu}X^{(4)}\mn\,,
\ea
with
\ba
\label{X4}
X^{(4)}\mn&\sim&24( \Pi^4\mn-\Pi \Pi^3\mn)+
12\mathcal{L}^{(2)}_{\rm der}\Pi\mn^2-4\mathcal{L}^{(3)}_{\rm der}\Pi\mn
+\mathcal{L}^{(4)}_{\rm der}\eta\mn\equiv 0\,,
\ea
with  $\mathcal{L}^{(2,3,4)}_{\rm der}$ given respectively in
\eqref{L2der}, \eqref{L3der} and \eqref{L4der}. The decoupling limit is
therefore well behaved up to the quintic order, and the number of free
parameters at higher orders suggests that one can always make appropriate
choices to avoid any ghost mode from appearing in the entire decoupling
limit. To be certain, one should however analyze a fully non-linear
theory, such as the one proposed in \cite{GG,Claudia}.

Motivated by the above obtained results,  we set up
in the next section  a general formalism for obtaining
the interactions to all orders.

Before we do so, some important comments are in order.
We might of course argue that the absence of the ghost up to the
quintic order represents no proof of the stability of the
theory even in the decoupling limit, since the ghost could be
pushed to the next order in interactions. It is also not a proof
of the consistency of the full theory, as was discussed in section 1,
since  the ghost may appear  away from the decoupling limit.
The arguments concerning these two
points, respectively,  are:

\begin{enumerate}
\item Beyond the quintic order, the number of free coefficients in the
interactions seems sufficient to eliminate pathological contributions of
the form $(\p\p \pi)^n$. Furthermore, beyond the quartic order all
conserved tensors of the form $X^{(n)}\mn \sim (\p \p \pi)^n\mn$ vanish
identically, and cannot lead to any ghost-like pathologies in the
mixing $h^{\mu\nu}X^{(n)}\mn$ between the helicity-0 and 2 modes.

\item  The ghost  may exist in a given order away from the decoupling limit
(say at the  quartic or higher order), but disappear  in the decoupling
limit.  If so,  then, the ghost should come with a mass greater than
$\Lambda_3$.  Then, the theory would be acceptable as an effective theory
below the  $\Lambda_3$ scale.  However,  at  scales above  $\Lambda_3$,
one would need to specify an infinite number of terms in the full nonlinear
theory in order to conclude whether or not the ghost is removed by
the resummation of these terms. This will be made more precise in the last section.

\end{enumerate}

\section{General formulation for an arbitrary order}

All our findings up to the quintic order presented in the previous section
can be formulated in a unified way, which may also suggest how things could
work at higher orders. For this, in the $N$th order expansion
(so far $N\le 5$), we introduce the notations
\ba
\label{SN}
\bar U_N(g,H)\equiv -
\frac{\mpl^2  m^2}{4} \sum_{i=2}^N \sqrt{-g}\, U_i(g,H) \,,
\ea
where the tensor $H\mn$ is defined as in section 2.
If the $N^{\rm th}$ order expression  for the function
$\bar U_{N}(g, H)$ satisfies
\beq
\bar U_{N}(g,H)\Big|_{h\mn =0,~A_\mu=0}  = {\rm total~derivative}\,,
\label{cond}
\eeq
(where $A_\mu$ denotes the helicity-1 field)
then, the decoupling limit Lagrangian for the helicity-0 and -2 interactions,
up to a total derivative,  takes the form:
\beq
{\cal L}^{\rm lim}_{\Lambda_3}=
-{1\over 2} { h}^{\mu\nu} {\cal E}_{\mu\nu}^{\alpha\beta}
{h}_{\alpha \beta}+
{h}^{\mu\nu}\bar X^{(N)}\mn(\pi)\,,
\label{conj}
\eeq
with the conserved tensor $\bar X^{(N)}\mn$:
\ba
\bar X^{(N)\, \mu\nu} (\pi)=
{\delta \bar U_{N}(g,H)   \over \delta h_{\mu\nu}}
\Big|_{h\mn =0,~A_\mu=0}\,.
\ea
We have checked that the above Lagrangian  gives rise to  equations of
motion with no more than two time derivatives and appropriate constraints
for $ N\le 5$. It seems reasonable to conjecture that this will also be
the case for $ N> 5$. Furthermore, in four dimensions $\bar X^{(N)}\mn$
can only contain a finite number of terms if it is local and conserved.
It is therefore likely that this formalism leads to a finite number of
interactions in the decoupling limit.

\vspace{0.1cm}

At a given order $n$ in the expansion,
there should be enough freedom to set the polynomial $U_n(g,H)$ appropriately,
so as to ensure that the leading interactions \eqref{Ln bad} enter as a total derivative of the form \eqref{Lder n},  or as $f(\Pi)\mathcal{L}^{(m)}_{\rm der}$ for $m\ge5$
and $f$ being an arbitrary function of $\Pi\mn$. The resulting leading
contribution is then of the form
\ba
\mathcal{L}^{(n)}=\frac{\beta}{\Lambda_3^{n-1}}h^{\mu\nu}X^{(n)}\mn\,,
\ea
where $\beta$ depends on the coefficient $c$'s, $d$'s, etc. and  $X^{(n)}\mn \sim \Pi^n\mn$ must be conserved as a straightforward consequence of reparametrization invariance in the decoupling limit (since higher interactions in $h$ are then suppressed). At each order $n$, there is a unique combination of $\Pi\mn^n$'s which is conserved. This combination is of the form
\ba
X^{(n)}\mn \propto \frac{\delta \mathcal{L}^{(n+1)}_{\rm der}}{\delta \Pi^{\mu\nu}}\,.
\ea
In four dimensions however, $\mathcal{L}^{(5)}_{\rm der}\equiv 0$ as pointed
out earlier, and the same remains true at higher orders. This further
implies that there is a  limit on the number of possible interactions
in the decoupling limit:
$X^{(n)}\mn\equiv 0$ for any $n\ge 4$.
This suggests that all theories of massive gravity
(with the scale $\Lambda_3$) can only have at most quartic couplings
between the helicity-0 and 2 modes in the decoupling limit.

\section{Massive gravity and the Galileon}

When making the generalized FP choice for the  coefficients
(\ref{c1 and c2}), (\ref{d1}-\ref{d4}), and (\ref {fs}),
the higher interactions in the decoupling limit
only arise as a coupling between the tensor mode
and the helicity-0 mode of the form
\ba
\mathcal{L}_{\rm int}={h}^{\mu\nu}\bar X^{(N)}\mn=h^{\mu\nu}
\(X^{(1)}\mn+\frac{1}{\Lambda_3^3}X^{(2)}\mn+\frac{1}
{\Lambda_3^6}X^{(3)}\mn\)\,,
\ea
where $X^{(1)}$ is given by \eqref{X1}, $X^{(2)}$ by \eqref{X2} and $X^{(3)}$
by \eqref{X3}. Moreover, as emphasized before,
$\partial^\mu X^{(i)}\mn=0$. We proceed further by noticing that
\ba
X^{(1,2)}\mn=\hat \E^{\alpha\beta}\mn Z^{(1,2)}_{\alpha\beta}\,,
\ea
with
\ba
Z^{(1)}\mn&=&\pi \eta\mn\,,\\
Z^{(2)}\mn&=&(6c_3-1)\p_\mu\pi \p_\nu\pi\,.
\ea
We can therefore diagonalize the action up to the
cubic order by performing  a local but nonlinear change of
the variable
\ba
h\mn=\hat h\mn+Z^{(1)}\mn+\frac{1}{\Lambda_3^3}Z^{(2)}\mn\,,
\ea
such that,  up to total derivatives, the Lagrangian  is
\ba
\label{galgen}
\mathcal{L}&=&-\frac 12 \hat h\mn \hat \E^{\mu\nu\alpha\beta}\hat h_{\alpha \beta} +\frac 32 \pi \Box \pi
+\frac32\frac{(6c_3-1)}{\Lambda_3^3}(\p \pi)^2 \Box \pi\\
&+&\frac{1}{\Lambda_3^6}\(\frac 12 (6c_3-1)^2-2(c_3+8d_5)\)(\p \pi)^2\([\Pi^2]-[\Pi]^2\)
+\frac{1}{\Lambda_3^6}\hat h^{\mu\nu}X^{(3)}\mn \nn\\
&-&\frac{5}{2\Lambda_3^9}(6c_3-1)(c_3+8d_5)(\p \pi)^2
\([\Pi]^3-3[\Pi][\Pi]^2+2[\Pi^3]\)
% \nn\\ &+&\ldots
\nn \,.
\ea
In the first line we see appearing the quadratic  and cubic
Galileon terms, \cite{Nicolis:2008in} (the usual kinetic term for $\pi$,
as well as the interaction present in DGP). In the second line we
notice the quartic Galileon interaction and finally the
quintic,  last interaction  of the Galileon family,  appears in the
last line.

By setting $c_3=-8d_5$ we precisely recover the Galileon family of terms up to quartic order,
and all the remaining couplings with the tensor mode disappear
at the quintic order. Since there is still a lot of freedom in the
coefficients at higher orders, it is only natural to expect this
result to be maintained to all orders.

On the other hand, if  $c_3\neq -8d_5$,
then  the last mixing term $h^{\mu\nu} X^{(3)}\mn$ does not seem to be
removable via any  {\it local} field redefinition.  This mixing term may
be crucial to address the issue of superluminality of  the massive theory,
as the Galileon without the mixing terms does exhibit superluminal
behavior \cite {Nicolis:2008in}.

In a more general case, as soon as the cubic Galileon is
present in (\ref{galgen}), we are also bound to have
either the  quartic Galileon and no other terms (for $c_3 =-8d_5$),
or a quartic mixing and the quintic Galileon (for $4(c_3+8d_5) = (6c_3-1)^2
\neq 0$),  or  all of the above terms together.

If however, the cubic Galileon is absent (for $c_3=1/6$),
one in general is left with the quartic Galileon and
the quartic mixing term.

Finally, notice also that for the specific choice $c_3=1/6$ and $d_5=-1/48$,
all the interactions at the scale $\Lambda_3$ disappear!  This may be an
example of a theory for which the decoupling limit picks up
a higher scale $\Lambda_\star > \Lambda_3$, if such a theory
exists. Alternatively, this may also be  a theory in which all the
nonlinear terms disappear in  the decoupling limit. This would
suggest that the theory has
no  strongly coupled behavior (\ie, no Vainshtein mechanism), and
would be ruled out observationally.

\section{Outlook}

The previous analysis shows that for appropriate choices of
interactions that generalize the Fierz-Pauli term to higher orders,
one can construct a consistent and local theory of massive gravity where
no ghost-like instabilities are present, at least up to the
quintic order in the
decoupling limit, and  positive prospects can be foreseen for higher
orders. In particular the connection with the Galileon generalization
of the cubic term appearing in the DGP decoupling limit provides a natural
framework for studying  ghost-free theories of gravity
\cite{Nicolis:2008in,deRham:2010eu}.

Furthermore, the decoupling limit considerations of this paper suggest
that the higher non-linear terms  in (\ref {PFS}-\ref{L5}) become
equally  important at the scale $\Lambda_3$. Since the scale
$\Lambda_3=(\mpl m^2)^{1/3}$ is very low
(typically $\Lambda_3\sim 10^{-9}$eV), the effective theory below
$\Lambda_3$ can only be used for large scale cosmological studies\footnote{
Once external classical sources, such as planets, stars, galaxies,..,
are present,
the energy scale of nonlinearities -- the Vainshtein scale -- depends on
the mass/energy of
the source and is significantly  lower \cite {DDGV}.}. To extend the scope of
applicability of massive gravity to shorter length scales, however, one would
need to go above  $\Lambda_3$, and, hence, the higher  interactions should  be
taken into account.  For a viable model, it will therefore be
necessary to consider all the higher polynomial interactions,  $U_n(g,H)$,
and not only the ones up to the quintic  order as presented here
(even though the decoupling limit may only have a finite
number of interactions).

A theory that provides such a resummation is the model of Refs.~\cite{GG,Claudia}. In particular, by integrating  out the auxiliary dimension in that model, one gets an infinite series of  interactions of the form (\ref {PFS}-\ref{L5}) and beyond,  with certain specific coefficients.  In \cite{cubic}, it has been checked that  the coefficients of the quadratic and  cubic terms were equal to those used in section 3 for the specific choice  $c_3=1/4$.
Thus,  in the decoupling limit, the theory is
ghost-free up to the cubic  order. Furthermore,  the
theory in the cubic order preserves  the Hamiltonian constraint even away from
the decoupling limit \cite {cubic}, and the BD term  cancels out in the
exact all-order Hamiltonian \cite {GG}.
Moreover, it was shown in Ref.~\cite{Claudia} that the nonlinear terms
giving rise to a ghost at a scale $\Lambda<\Lambda_3$ cancel out in that
specific theory. These findings constitute an important evidence (but not a proof yet)
that the theory  of \cite{GG,Claudia} may be consistent, at least classically,
to all orders.

How about other possible theories of massive gravity that would yield
the terms discussed here with the coefficients still consistent with
the absence of the ghost,  but not coinciding with the ones obtained
in \cite {cubic}?  Is there any hope for these
theories away from the decoupling limit and above the scale $\Lambda_3$?
Naively, the answer seems to be a negative one:
As was shown in \cite{Creminelli}, in the
order-by-order expansion,  and beginning with the quartic order,
one cannot avoid higher powers of the lapse function in
the Hamiltonian, and hence, the emergence of the sixth  degree of freedom
(which typically is a ghost) seems to be unavoidable in
massive gravity \cite{Creminelli}.

However, there may  be a way to circumvent this problem in the full
theory  if its  Hamiltonian, due to a resummation of perturbative terms,
ends up having a very special dependence on the  lapse and
shift functions. Here we demonstrate this in a toy example,
that is motivated by the Hamiltonian of the theory
\cite{GG,Claudia} discussed in \cite {GG}.

Consider the toy Hamiltonian:
\beq
H= N\left( R^0 + m^2 f(\gamma) \right ) + N_j\left(
R^j +m^2 Q^j(\gamma)\right)+ m^2 P(\gamma) {N_j
N^j\over 2N}\,,
\label{H}
\eeq
where $N,N_j,\gamma_{ij}$,  and $R^0, R^j$, are the standard ADM  variables and functions respectively \cite{ADM}; $f(\gamma),Q_j(\gamma) $ and $P(\gamma)$ are some  functions that modify the GR constraints by the mass terms. The shift function $N_j$ is not a Lagrange multiplier, but is algebraically determined, as it should be the case for a massive theory with five degrees of freedom. However, the lapse functions also enters in the last term in a way  that seems  to prevent it to be a Lagrange multiplier, and if so, it would give rise to the sixth degree of freedom.  This is not the case, however:  One can introduce a new variable
$n_j \equiv N_j/N$ in terms of which the Hamiltonian  reads
\beq
H= N(R^0 + m^2 f(\gamma)) + N n_j( R^j +m^2 Q^j(\gamma))+
N m^2 P(\gamma) {n_j n^j\over 2}\,.
\label{H1}
\eeq
The shift $n_j$,  still has no conjugate momentum,
hence  $\delta H /\delta n_j =0$. This determines the new shift variable,
$n^j =- (R^j +m^2 Q^j(\gamma))/(m^2P(\gamma))$, and yields the following
Hamiltonian
\beq
H|_{n_j}=
N\left ( R^0 + m^2 f(\gamma)  - { (R_j +m^2 Q_j(\gamma))^2
\over 2 m^2 P(\gamma) }
\right )\,.
\label{H3}
\eeq
Here, the lapse does certainly appear as the Lagrange multiplier.
Hence, the BD term does not arise, and the theory does not propagate the sixth
degree of freedom\footnote{In general,  it could still be propagating ``$5.5$''
modes even if the Hamiltonian constraint is maintained.
For instance, since the toy model described by (\ref {H}) is not Lorentz
invariant  for general functions $f$, $Q_j$ and $P$,  there may exist
non-propagating instantaneous
modes in this model. For discussions of  related issues see,
\cite{GGLuca}.  In contrast, the model of Refs.~\cite{GG,Claudia} is
4D Lorentz-invariant and the instantaneous mode in 4D is not expected. For
a rigorous proof
that there are only 5 degrees of freedom, and not ``$5.5$'', however,
a detailed study of the algebra of the Hamiltonian constraint should be
performed. The fact that the
decoupling limit gives only 5 degrees of freedom is an important
hint that the full theory is not likely to have the  extra ``$0.5$''
degree of freedom.}.

On the other hand, a direct perturbative expansion of the last
term in (\ref {H})
in powers of $\delta N = N-1$ with subsequent truncation of this series at
any finite nonlinear order, necessarily yields higher powers of  $\delta N$
in the Hamiltonian\footnote{Note that away from the decoupling limit, and
at a nonlinear order,  $\delta N$ is the right variable and not
$h_{00} = 1-N^2 +N_j^2$, which was used before as the lapse
in the decoupling limit.}. Naively, this truncated theory would give
rise to the potentially false impression that the lapse is not a Lagrange
multiplier, and that there is a sixth degree of freedom in the model.

Noticing  that  the higher powers of   $\delta N$   at any finite nonlinear
order emerge  from the expansion of the theory (\ref{H}) is trivial, in this toy model.
However, a similar, albeit more complicated structure, emerges in the Hamiltonian
of the model of \cite{GG,Claudia} (see \cite{GG}) and  the fact that the terms
in the expansion come up from a single term in the exact Hamiltonian is not as
simple to observe.

\vspace{0.1cm}

Last, but not least, in this work we discussed
the classical theory. Generic  quantum loop corrections are expected
to renormalize and detune the coefficients of the polynomial terms
needed to avoid the ghost. One way to be protected against this
problem is to have a theory in which the tuned coefficients
automatically emerge as a consequence of a symmetry that would be
respected by the loop corrections.  In this respect, the recent
findings of \cite {cubic} that the cubic terms with the automatically tuned
coefficients emerge as an expansion of the  theory, which by itself
exhibits an evidence for a hidden nonlinearly realized symmetry,
makes us hopeful for the existence of a quantum-mechanically
stable effective field theory of massive gravity.

\section*{\small Acknowledgments}

We would like to thank Cedric Deffayet, Gia Dvali, Massimo Porrati,
Oriol Pujolas,
Andrew Tolley and Filippo Vernizzi for useful comments. The work of
GG was supported by NSF grant PHY-0758032, and that of CdR by the SNF.

\appendix

\renewcommand{\thesection}{Appendix \Alph{section}}

\section{Decoupling limit with the opposite sign in $H\mn$}
\label{AppCreminelli}

As mentioned in section \ref{GI}, the expression \eqref{Hmn} for $H\mn$
differs by  a minus sign in front of the third term on the r.h.s.
from its counterpart considered in Eq.~(5) of \cite{Creminelli}:
\ba
\label{HmnCreminelli}
H\mn=\frac{h\mn}{\mpl}+\partial_\mu \pi_\nu + \partial_\nu \pi_\mu + \eta_{\alpha\beta}\partial_\mu \pi^\alpha \partial_\nu \pi^\beta\,,
\ea
To emphasize the importance of this sign difference, we show that we recover the results of  Ref. \cite{Creminelli} when deriving the decoupling limit using \eqref{HmnCreminelli}, but stress that the Bianchi identity is then not
satisfied, as a consequence of the fact that $H\mn$ is then not
a covariant tensor if $g\mn$ and $h\mn$ are conventionally defined.

Up to the cubic order, the Lagrangian in the decoupling limit is then
\ba
\tilde{\mathcal{L}}&=&-\frac 12 h^{\mu \nu}
\hat \E^{\alpha\beta}_{\mu\nu}h_{\alpha\beta}+
 h^{\mu \nu} \tilde X^{(1)}\mn\\
&&-\frac{1}{4\Lambda_5^5}\((8c_1+4)
[\Pi^3]+(8c_2-4)[\Pi][\Pi^2]+8c_3[\Pi]^3\)\nn
+\frac{1}{\Lambda_3^3}h^{\mu\nu}\tilde X^{(2)}\mn\,,
\ea
with  $\tilde X^{(1)}\mn= X^{(1)}\mn$, since both approaches only
differ at quadratic order in $\pi$,  and
\ba
\tilde X^{(2)}\mn=-\(3c_1-\frac 32\)\Pi^2\mn-2(1+c_2)[\Pi]\Pi\mn
+\(\frac 12 -3c_3\)[\Pi]^2\eta\mn-c_2 [\Pi^2]\eta\mn\,.
\ea
Setting $c_1=2c_3-\frac 1 2$ and $c_2=-3c_3+\frac 12$ to obtain the total
derivative combination \eqref{L3der}, we get
\ba
\label{X22}
\tilde X^{(2)}\mn=-6(c_3-\frac 12 )
(\Pi^2\mn-[\Pi]\Pi\mn)-(3c_3-\frac12)\([\Pi]^2-[\Pi^2]\)\eta\mn\,,
\ea
which is not conserved for any choice of $c_3$ since the reparametrization
invariance is not present with this choice of $H\mn$,
and the Bianchi identity  has no reason to be satisfied.

Similarly at the quartic order, we would need to impose the
relation between the coefficients
$d_1=-6d_5-\frac{1}{16}(24c_3-5)$,
$d_2=8d_5+\frac{1}{4}(6c_3-1)$,
$d_3=3d_5+\frac{1}{16}(12c_3-1)$, and
$d_4=-6d_5-\frac34 c_3$, to cancel the terms of the
form $\Lambda_4^{-8}(\p \p \pi)^4$.
The mixing with the helicity-2 mode, will then enter with the
quantity $\tilde X^{(3)}\mn$ as derived in \cite{Creminelli}:
\ba
\tilde X^{(3)}\mn &=&(-1+9c_3+24d_5)(\Pi^3\mn-[\Pi]\Pi^2\mn)
- (9c_3+24d_5)\Pi\mn ([\Pi^2]-[\Pi]^2)\\
&&-(c_3+8d_5)\eta\mn\([\Pi]^3-3[\Pi][\Pi^2]+2[\Pi^3]\)\nn
\,.
\ea
As noticed in \cite{Creminelli}, not only there
would then be no choice of $c_3$ and $d_5$ for which this interaction
disappears, and it would always lead to higher derivative equations of
motion, suggesting a ghost-like instability. However the fact that
$\tilde X^{(3)}\mn$ is not conserved is an artifact of
the sign  choice in the expression for  $H\mn$, that does not lead to
reparametrization invariant results.

\end{document}